\documentclass[showpacs,preprint,prl]{revtex4}

\usepackage{graphicx}

\begin{document}

\title{Overcharging, charge inversion and reentrant condensation: Using highly-charged 
polyelectrolytes in tetravalent salt solutions as an example of study} 
\author{Pai-Yi Hsiao}
\email[E-mail: ]{pyhsiao@ess.nthu.edu.tw}
\affiliation{%
    Department of Engineering and System Science, 
    National Tsing Hua University, 
    Hsinchu, Taiwan 30013, R.O.C.
}

\date{\today}

\begin{abstract}
We study salt-induced charge overcompensation and charge inversion of flexible polyelectrolytes 
via computer simulations and demonstrate the importance of ion excluded volume. Reentrant 
condensation takes place when the ion size is comparable 
to monomer size, and happens in a middle region of salt concentration. In a high-salt region, 
ions can overcharge a chain near its surface and charge distribution around
a chain displays an oscillatory behavior. Unambiguous evidence obtained by 
electrophoresis shows that charge inversion does not necessarily appear with 
overcharging and occurs when the ion size is not big. These findings suggest 
a disconnection of resolubilization of polyelectrolyte condensates at
high salt concentration with charge inversion.
\end{abstract}


\maketitle

Overcharging is a phenomenon, happening in many chemical and biological systems. 
It accounts for the situation that more counterions are attracted to the surface of a 
charged macromolecule in an electrolyte solution than needed to neutralize the bare 
surface charge~\cite{overcharging_review}. 
Recent studies show that counterions can form strongly correlated liquid on a charged 
surface and the negative chemical potential of the liquid provides an additional 
attraction of the counterions to the surface, leading to overcharging~\cite{shklovskii}.  
There is a second phenomenon, closely related to overcharging,  in which the effective 
charge of a macromolecule inverts its sign in an electrolyte solution under certain 
condition, resulting in the reversal of electrophoretic mobility~\cite{reversal_exp}.
We will call this phenomenon ``charge inversion''~\cite{note-definition}. 
In charge inversion, all the ions bound to the molecule are considered because they
contribute to the effective charge, whereas in overcharging, only the  charge 
overcompensation near a molecular surface is discussed. 
Without doubt, overcharging occurs while charge inversion takes place. However, we do not
know for sure if the occurrence of the previous inevitably brings in the latter. This is
a fundamental question addressed in this study.

In the past decades, there is a resurgent interest in condensing DNA 
using charged small molecules or salts for diverse applications~\cite{bloomfield96}. 
This kind of study deals with highly-charged polyelectrolytes (PE) and multivalent 
counterions, which goes beyond the description of the classical mean-field theory based 
on Poisson--Boltzmann equation.  Upon addition of multivalent salt, 
PE is firstly condensed from a solution, due to ionic screening or bridging;  
the condensed PE is then dissolved into the solution 
when salt concentration is increased to a high value~\cite{reentrant_exp}. 
Nguyen \textit{et al.} recently  explained these phenomena, called reentrant condensation, 
using the idea of overcharging~\cite{nguyen00}.  
When the surface charge of PE is reduced to nearly zero by condensed multivalent 
counterions,  the correlation induced short range attraction dominates the Coulomb 
repulsion, leading to PE condensation. 
At a more elevated salt concentration, PE can get overcharged and the 
Coulomb repulsion dominates; the condensed PE thus reenters into the solution. 
In their theory, overcharging implies the sign inversion of the effective charge, 
and consequently, the reversal of the moving direction of PE in an electric field was 
expected. Reversal of electrophoretic mobility has been observed since decades  
ago~\cite{reversal_exp} and firstly explained by theorists in 1980's~\cite{gonzales-tovar85}.  
Nonetheless, the connection with the redissolution of PE
is still not well understood.  Recent study showed that the resolubilization of fd virus 
bundles happened at high salt concentration without mobility reversal~\cite{wen04}. 
Controversy hence arises.  
Integral equation theories~\cite{gonzales-tovar85,Lazada-Cassou99,quesada-perez05} and 
simulations~\cite{deserno01,tanaka,hsiao} have both confirmed the 
possibility of charge overcompensation next to a PE surface. 
They have also demonstrated that the ion size is an important factor to affect the properties of
a macromolecule. Moreover, an oscillatory behavior in the integrated charge distribution
around a molecule has been found in these studies, which indicates the formation
of ionic layers, alternating the sign of charge.  It therefore becomes unclear whether the
net charge of the whole PE-ion complex inverts its sign or not, even if there is 
overcharging in the vicinity of the surface. 
Noticeably, theorists also predicted the importance of Bjerrum association and 
the dependence of charge inversion on the ion size~\cite{solis01}.

The main difficulty in calculation of the net charge comes from 
no precise information about the border of a macroion-ion complex.     
A simple but accurate way to circumvent this difficulty is to 
do electrophoresis in a weak electric field.  In this Letter, we use multiple 
PEs in tetravalent salt solutions as an example and computer simulation as a tool 
to investigate the relationship between overcharging, charge inversion, and
redissolution of PE condensation. 
Our system contains four bead--spring chains and many charged spheres, which model 
anionic polyions and monovalent cations, dissociated from the PEs, and tetravalent
cations and monovalent anions, dissociated from the salts.  Each chain consists of 
48 monomers and each monomer carries a negative unit charge $-e$.
The excluded volume of monomers and ions is modeled by a shifted Lennard-Jones potential, 
$U_{\rm LJ}(r)= \varepsilon_{\rm LJ} [2(\sigma/r)^6-1]^2$,
truncated at the minimum, with coupling strength
$\varepsilon_{\rm LJ}=k_{\rm B}T/1.2$,
where $k_{\rm B}$ denotes Boltzmann's constant and $T$ the
absolute temperature.  We assume that the monomers, monovalent cations and anions
have equal effective diameter $\sigma=\sigma_{m}$ and vary 
the effective diameter of tetravalent cation $\sigma_{t}$ from 
$0$ to $4.0\sigma_{m}$.  This size variation covers a broad range of interest for 
theorists and experimentalists, discussing from point charges to large charged colloids. 
The virtual springs jointing monomers on a chain are 
described by the FENE potential~\cite{stevens95},
$U_{\rm FENE}(b)= -0.5 k b^2_{max} \ln (1-b^2/b^2_{max})$,
with maximum extension $b_{max}=2\sigma_{m}$ and
spring constant $k=7\varepsilon_{\rm LJ}/\sigma_{m}^2$.
Solvent is simulated as a medium of uniform dielectric constant~$\varepsilon$, 
and hence, the Coulomb interaction between two particles 
of valences $z_i$ and $z_j$ is equal to $k_{\rm B}T\lambda_{\rm B} z_i z_j/r$
where $\lambda_{\rm B} \equiv e^2/(4\pi\varepsilon\varepsilon_0 k_{\rm B}T)$
is the Bjerrum length and $\varepsilon_0$ the vacuum permittivity.
We set $\lambda_{\rm B}=3\sigma_m$. The model is representative of a prototypical PE 
such as sodium polystyrene sulfonate in an aqueous solution at room temperature. 
The system is placed in a periodic cubic box and subject to a uniform external 
electric field pointing to $+\hat{x}$ direction.
Langevin dynamics simulations are employed in this 
study~\cite{note-lammps} and the equation of motion reads as
$ m \ddot{\vec{r}_i} = -m\gamma_i \dot{\vec{r}_i} + \vec{F}^{(c)}_i +
z_i e E \hat{x} + \vec{{\eta}}_i(t)$
where the particles are assumed to have  identical mass $m$ and  
$m\gamma_i$ is the friction coefficient.
According to the Stokes law, $\gamma_i$ is linearly proportional 
to the particle size $\sigma_{i}$;
we thus choose $\gamma_i=(\sigma_i/\sigma_{m})\tau^{-1}$ 
with $\tau=\sigma_{m} \sqrt{m/(k_{\rm B}T)}$.
$\vec{F}^{(c)}_i$  is the conservative force acting on $i$, and $\vec{\eta}_i$ is 
the white noise satisfying the fluctuation-dissipation
theorem.  $E$ is the strength of the electric 
field and we set $E=0.05 k_{\rm B}T/(e\sigma_{m})$. The monomer concentration 
is fixed at $C_m=0.008 \sigma_{m}^{-3}$. 
This $C_m$ is well below the overlap threshold and the results
have been shown representative for the range 
$0.00032 \sigma_{m}^{-3} \leq C_m \leq 0.016 \sigma_{m}^{-3}$~\cite{hsiao}.
In the rest of the text, we use $\sigma_{m}$, $\tau$, $e$, and $k_{\rm B}T$
as the units of length, time, charge, and energy, respectively.
Therefore, concentration will be measured in unit $\sigma_{m}^{-3}$,
strength of electric field in unit $k_{\rm B}T/(e\sigma_{m})$,
and so on.

\begin{figure}[htbp] 
\begin{center}
  \includegraphics[width=0.6\textwidth,angle=270]{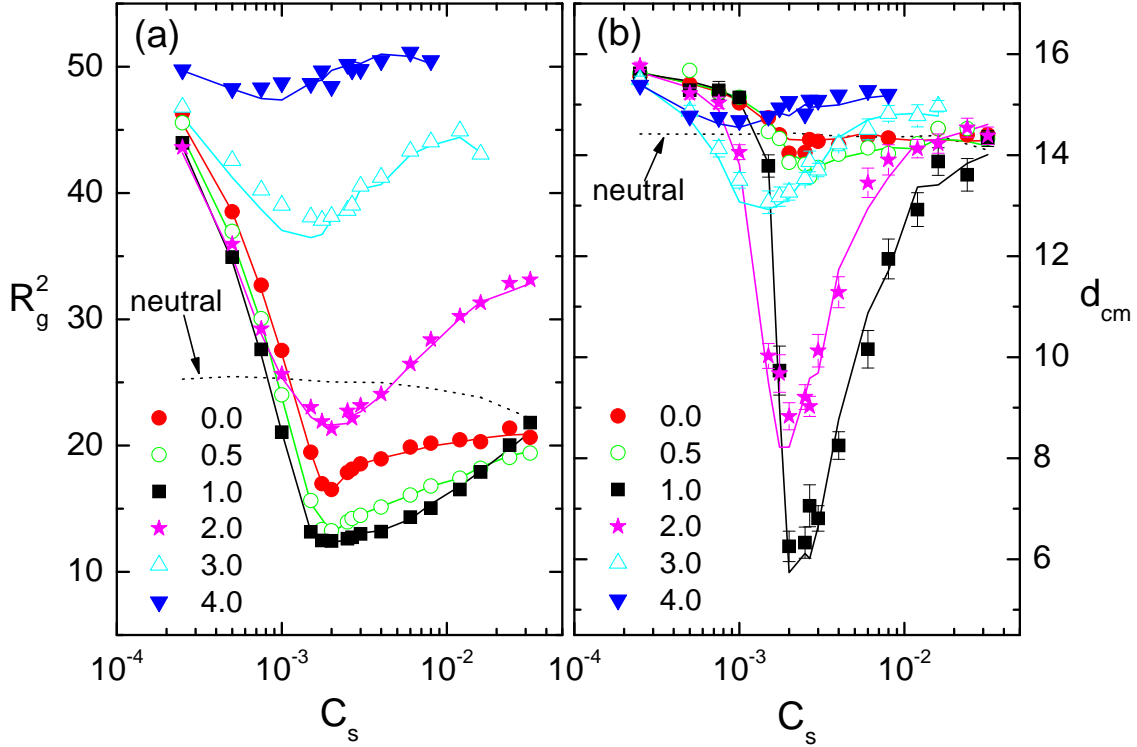}
  \caption{(a) $R_g^2$ and (b) $d_{cm}$ as a function $C_s$ for different 
  values of $\sigma_t$ indicated in the figures.
  The symbols denote the data obtained in the presence of the electric field, 
  whereas the solid curves nearby are the ones obtained in the absence of it.
  The dotted curve is the data for neutral polymers.}
  \label{fig:Rg2&CMdist}
\end{center}
\end{figure}

We first verified that $E=0.05$ is a weak field for our system so that    
the conformation of the chains and the structure of the solution
are almost not modified by the electric field.  
It is justified by comparing both 
the mean square radius of gyration $R_g^2$ and 
the mean center-of-mass distance between chains $d_{cm}$
in the presence and in the absence of the electric field,  
shown in Figs.~\ref{fig:Rg2&CMdist}(a) and \ref{fig:Rg2&CMdist}(b).
The difference between the two set of data is very small. 
We have verified that the electric field applied here is much weaker 
than needed to deform the chains. 

In Fig.~\ref{fig:Rg2&CMdist}(a), we observed that the chain size
for $\sigma_t<3.0$ decreases with increasing $C_s$ up to 
$C_s^*(=0.002)$ at which the total tetravalent cations and
the 4 PE chains are at charge equivalence.
A striking phenomenon occurs once $C_s$ is increased beyond $C_s^*$:
$R_g^2$ increases or in other words, the chains swell.
This chain collapsing and swelling can be regarded as a single-chain 
version of the macroscopic phase separation and redissolution 
of PEs happening in multivalent salt solution~\cite{reentrant_exp},
and have been reported in our previous study~\cite{hsiao}. 
For the cases with larger $\sigma_t$, the collapsing and swelling 
basically do not take place. It demonstrates a strong influence 
of ion excluded volume on the chain size. 
In addition to the variation of single-chain size, multivalent salt 
can also provoke subsequently multi-chain aggregation and 
segregation. 
When the ion size is intermediate ($\sigma_t=1.0$ and $2.0$), $d_{cm}$ curve 
in Fig.~\ref{fig:Rg2&CMdist}(b) shows a deep valley near $C_s^*$, with 
its value smaller than $2R_g$, which strongly indicates chain aggregation. 
At high $C_{s}$, $d_{cm}$ increases to the value for neutral polymers; 
PEs hence segregate. It is worth noticing that even there occurs 
single-chain collapsing in the region $C_s \le C_s^*$, multichain aggregation 
may not take place, for example, when $\sigma_t=0.0$ and 0.5.
 
\begin{figure}[htbp] 
\begin{center}
  \includegraphics[width=0.6\textwidth,angle=270]{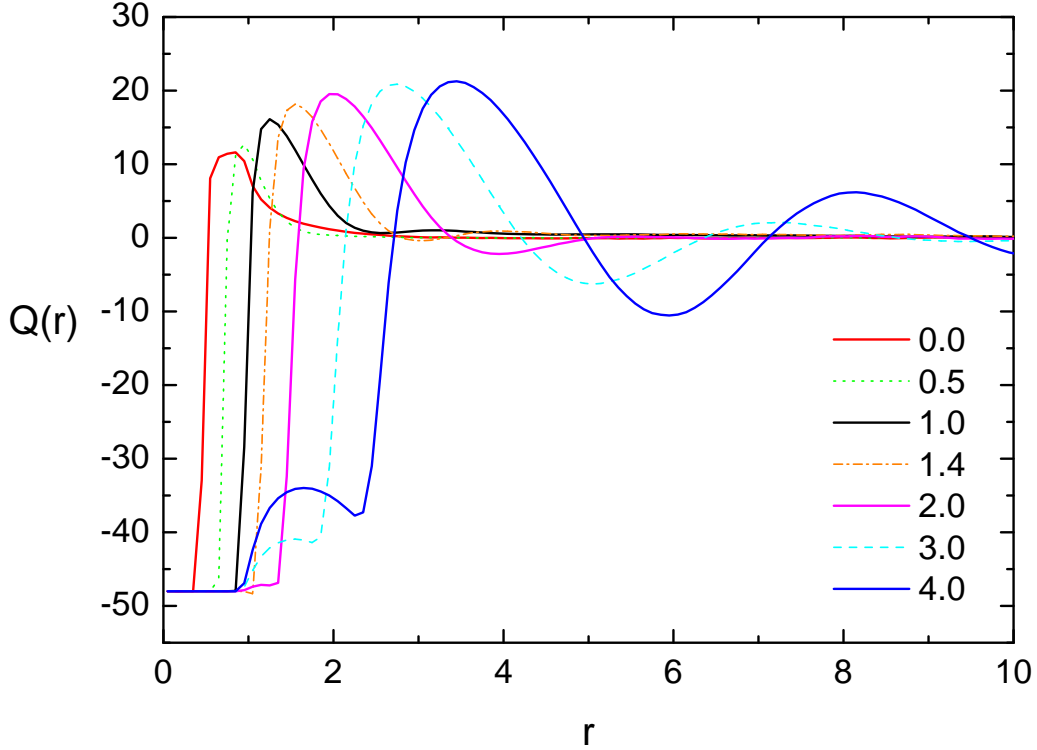}
  \caption{$Q(r)$ at $C_{s}=0.008$ for different values of $\sigma_t$ 
   indicated in the figure}
  \label{fig:IonDistP4N48e005}
\end{center}
\end{figure}

We then studied the integrated charge distribution $Q(r)$ around a PE.
$Q(r)$ denotes the total charge inside a \textit{wormlike} tube 
which is the union of the spheres of radius $r$ centered at each monomer 
center of a PE.  For $C_{s}<C_{s}^*$, $Q(r)$ increases monotonically 
from $-48$ to $0$ with $r$, due to electroneutrality~\cite{hsiao}. 
If $C_{s}>C_{s}^*$, a positive peak appears in $Q(r)$ next to the PE surface. 
This peak results mainly from an excess number of the tetravalent counterions 
appearing near the surface; the chains are thus overcharged. 
The degree of overcharging increases with $C_{s}$. At high $C_{s}$, $Q(r)$ 
eventually shows an oscillatory behavior around zero, which suggests a repetitive
overcompensation of charge inside a wormlike tube. Fig.~\ref{fig:IonDistP4N48e005} shows an 
example of $Q(r)$ at high $C_{s}=0.008$ for different $\sigma_t$.
We observed that the positive peak near the surface and the oscillatory
behavior of $Q(r)$ are both enhanced as $\sigma_t$ is increased.  
These phenomena could be explained from the point of view of entropy~\cite{messina02}. 
Increasing ion size decreases the free moving space of ions and thus,
decreases the entropy of the solution. The correlation of particles therefore 
becomes stronger, manifested, for instance, in $Q(r)$ curve with a more 
pronounced peak and oscillatory behavior.
Notice that for large $\sigma_t$, monovalent cations can become energetically
competitive with tetravalent ones, to condense onto a PE.
It is the condensation of monovalent cations which leads to the small bumps at 
$r=1.5$ for $\sigma_t=3.0$ and 4.0.

\begin{figure}[htbp] 
\begin{center}
  \includegraphics[width=0.6\textwidth,angle=270]{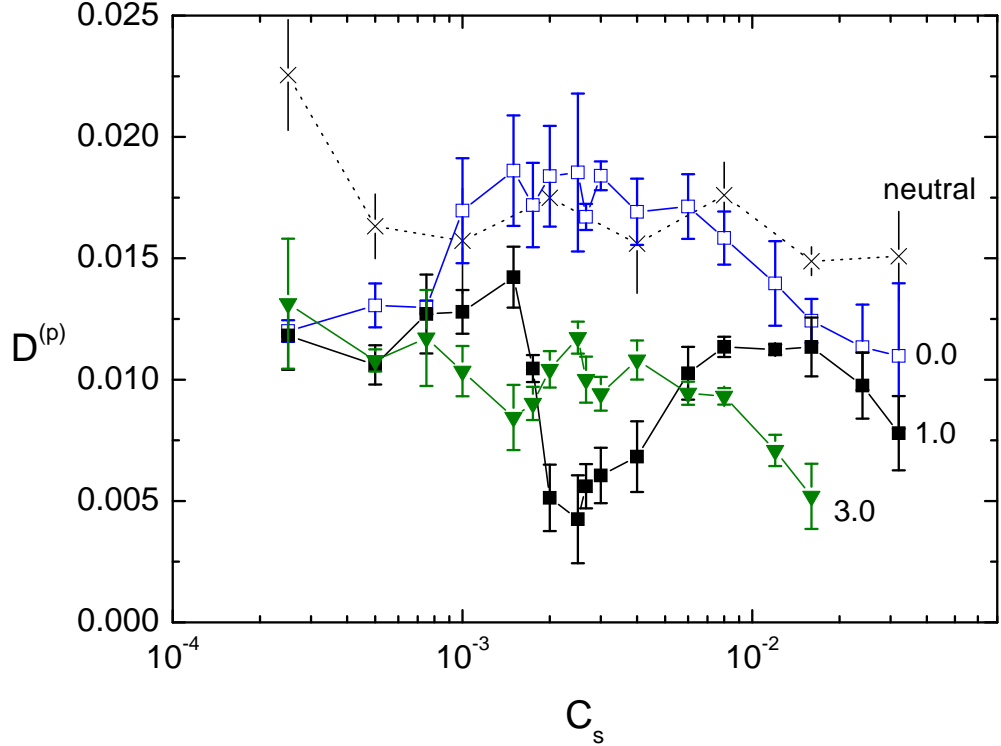}
  \caption{$D^{(p)}$ as a function of $C_s$ for $\sigma_t=0.0$
  1.0 and 3.0. The dotted curve denotes the data for neutral polymers.  }
  \label{fig:D_Cs}
\end{center}
\end{figure}

The net charge of a PE-ion complex can be calculated by 
$Q_{\rm eff}=v^{(c)}_{d}/(\mu^{(c)} E)$, provided that   
the drift velocity $v^{(c)}_{d}$ in the electric field $E$
and the kinetic mobility $\mu^{(c)}$ of the complex are known. 
Since PE chains move with complexes, we approximated 
$\mu^{(c)}$ by chain mobility $\mu^{(p)}$, calculated using 
Einstein relation: $\mu^{(p)}=D^{(p)}/(k_B T)$. 
$D^{(p)}$ is the diffusion coefficient obtained 
by taking time derivative of the mean square displacement 
of the center of mass of a PE chain in $\hat{y}$- and 
$\hat{z}$-directions and the results are shown in 
Fig.~\ref{fig:D_Cs}. 
We found that the referenced $D^{(p)}$ for neutral polymers decreases with 
increasing $C_{s}$ due to the jamming effect of the salt ions presented in solutions. 
For PEs, $D^{(p)}$ is smaller than the referenced $D^{(p)}$ while $C_s<0.001$; 
it is because extra friction forces act on the condensed counterions,  
which diminishes the chain diffusivity. Since the majority of the condensed ions 
is monovalent in this salt region, the diffusivity is not sensitive to the 
size of the tetravalent counterions.  
Differently, in the mid-salt region around $C_s^*$, the main condensed 
ions are tetravalent; $D^{(p)}$ depends sensitively on $\sigma_t$.
For $\sigma_t=0.0$,  there is no dragging force acting on the tetravalent ions;
$D^{(p)}$ is hence large and the curve shows a hump structure. 
For $\sigma_t=1.0$, the dragging force takes effect and in addition, 
chains aggregate and form big complexes, which leads to a drastic decrease of 
the diffusivity.  
The importance of the latter effect can be seen from the absence of the 
drastic decrease in $D^{(p)}$ for $\sigma_t=3.0$ since 
the chain aggregation does not take place. 
In the high-salt region, the chains segregate for any $\sigma_t$.
The decrease of $D^{(p)}$ is a result of the combination of the two effects: 
ionic jamming and PE complexation with large number of ions.  

\begin{figure}[htbp] 
\begin{center}
  \includegraphics[width=0.6\textwidth,angle=270]{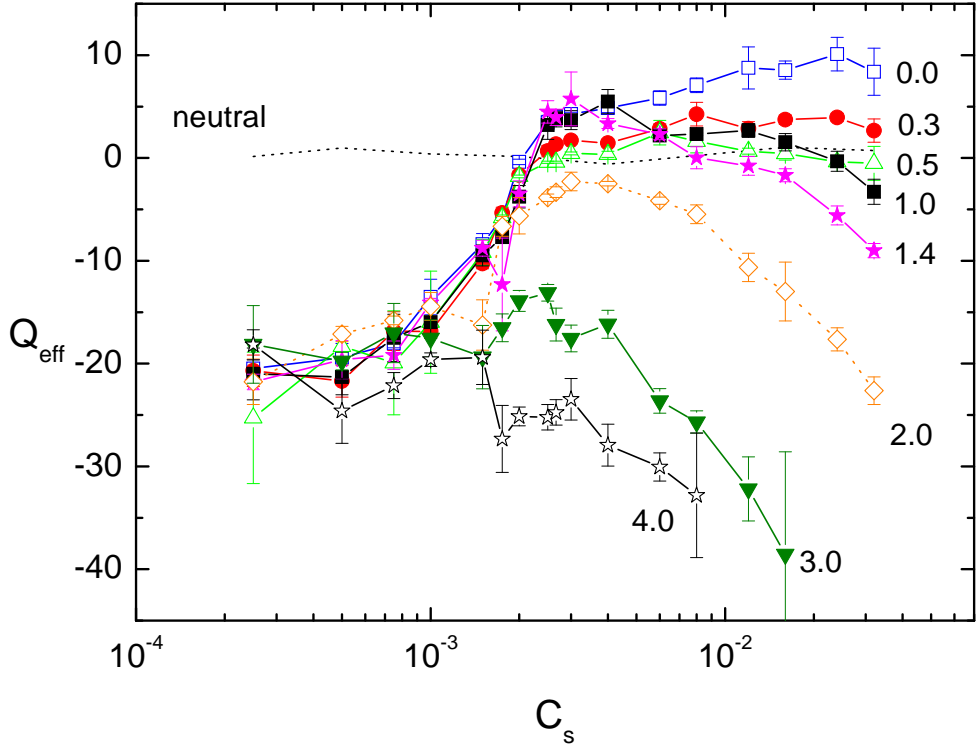}
  \caption{$Q_{\rm eff}$ as a function of $C_s$ for different
   $\sigma_t$. The value of $\sigma_t$ is given near the associated curve. 
   The dotted curve denotes the data for neutral polymers.}
  \label{fig:Qeff_Cs}
\end{center}
\end{figure}

We also approximate $v^{(c)}_{d}$ by the chain drift velocity $v^{(p)}_{d}$.
The computed $Q_{\rm eff}$ are presented in Fig.~\ref{fig:Qeff_Cs}.
In the low-salt region $C_s<0.001$, $Q_{\rm eff}$ weakly depends on 
the ion size and attains a value roughly equal to $-20$. 
This value is smaller than the prediction of Manning condensation 
theory~\cite{manning69}, which states a reduction of the effective 
line charge density on an infinite rodlike chain to $-e/\lambda_B$ 
and yields an effective chain charge $-17.2$ for our 
case~\cite{note-manning} regardless that our PEs are finite and flexible.
$Q_{\rm eff}$ turns to become sensitive to $\sigma_t$ 
while $C_s$ is larger than 0.001. 
For $\sigma_t<2.0$, $Q_{\rm eff}$ increases and 
surpasses zero around $C_{s}^{*}$.
It clearly demonstrates the occurrence of charge inversion.
$Q_{\rm eff}$ attains a value of roughly $10$ at high $C_{s}$ for point
tetravalent ions and the value gradually decreases as $\sigma_t$ is 
increased. For $\sigma_t=1.0$ and $1.4$, a hump appears in $Q_{\rm eff}$ 
curve and crosses zero line twice, indicating that charge inversion  
takes place only in a window of salt concentration.
For $\sigma_t\ge 2.0$, although showing a hump, the curve stays 
completely in the negative region --- no charge inversion takes place.   
$Q_{\rm eff}$ at high $C_s$ can be even more negative than in a salt-free solution
for large ion size such as $\sigma_t=3.0$ and 4.0. 
We remark that $Q_{\rm eff}$ obtained here stands for the mean net 
charge of a PE complex which may contain multiple chains, 
and thus, should not be simply interpreted as the net charge of a single PE, 
specially in the mid-salt region where chains aggregate. 
A general behavior of decreasing of $Q_{\rm eff}$ in the high-salt region 
was found.  The onset of the associated phenomenon, the decrease of the 
electrophoretic mobility at high $C_s$, has been observed in 
experiments~\cite{ottewill68,quesada-perez05} but never been emphasized before.  Both 
mean field theory and integral equation theory fail to predict this decrease.  
Noticeably, from a comparison between Fig.~\ref{fig:IonDistP4N48e005} and 
Fig.~\ref{fig:Qeff_Cs}, we can see that the location of the slipping plane of 
a PE in the notion of electrophoresis depends on the ion size. 
The smaller the ion size, the closer the plane to a PE. 
The location is roughly at $r=(\sigma_m+\sigma_t)/2$. 

We remark that the hydrodynamic interaction (HI) was not taken into account in this study. 
Theorists have shown that HI is screened in a typical condition of electrophoresis and 
thus weak compared to the electrostatic binding force~\cite{viovy}. 
This prediction has been confirmed by simulations~\cite{tanaka}.
In our case, the applied field is weak and the Debye screening length is short.  
Therefore, HI is negligible. 
Our results show that even if overcharging happens next to a 
PE surface in a high-salt region (Fig.~\ref{fig:IonDistP4N48e005}), 
charge inversion does not necessarily come with it. 
Moreover, a disconnection of the reentrant transition with the charge inversion was observed:
chains segregate at high $C_s$ for $\sigma_t=2.0$ (Fig.~\ref{fig:Rg2&CMdist}(b))
but charge inversion does not occur. 
We have specially performed simulations for $\sigma_t=2.0$ in several electric fields  
of another order of magnitude, ranging from $E=0.005$ to $0.1$, and found consistent 
electrophoretic mobility and ion distributions around a PE. 
It confirms that the applied field is weak enough so that the following scenario 
does not happen: PEs are stripped of their condensed counterions by the electric field, 
which leads to the negative $Q_{\rm eff}$. 
Even though there is a strong segregation for the case $\sigma_t=1.0$,
the absolute value of $Q_{\rm eff}$ is small.  
These findings suggest that chain redissolution does not come from Coulomb 
repulsion owing to the inversion of the net charge to a non-negligible value.
Moreover, the reversal of electrophoretic mobility can take place only 
when the size of multivalent counterions is small, complying with
the theory of Nguyen \textit{et al.}~\cite{nguyen00} in which they assumed point ions. 
Solis and Olvera de la Cruz~\cite{solis01} predicted a different result:  
no charge inversion for small ions.  The difference comes from their assumption that  
the ions have identical size. It artificially aggrandizes the effect of ion 
association when the ions are small.  Simulations using a model similar to theirs 
(cf.~the second paper in Ref.~\cite{hsiao}) gave consistent results with their prediction.

We propose the following mechanism for chain decondensation. If $C_s$ is high, the increase 
of energy related to the segregation of chains will be small because tetravalent counterions 
present abundantly 
in the bulk solution and interact with the segregated chains, which reduces the energy.  
Therefore, the entropy increase related to segregation can become a dominated
term, which lowers the free energy. Consequently, the system favors chain segregation. 
The effective charge is another issue. If $\sigma_t$ is large, the electrostatic 
correlation will be small.  The entropy effect related to the ion excluded volume does 
not produce sufficiently strong complexation to lead to charge inversion, although the 
local overcharging can still be possible. 
 
In summary, we have studied ``overcharging'' and ``charge inversion'' of PEs in 
tetravalent salt solutions. The crucial role of the ion excluded volume on the PE 
properties has been demonstrated.  Overcharging does not necessarily bring in charge inversion
and the effective charge is a nonmonotonic function of salt 
concentration. Unambiguous evidence has been presented that chain redissolution 
induced by tetravalent salt can take place without charge inversion. 

\begin{acknowledgments}
  This material is based upon work supported by the National Science Council,
  the Republic of China, under Grant No.~NSC 95-2112-M-007-025-MY2.
  Computing resources are supported by the National Center for High-performance 
  Computing.
\end{acknowledgments}


\begin{thebibliography}{26}
\bibitem[{\citenamefont{note3}(2008)}]{note-manning}
\bibinfo{note}{Our PE chain has 47 bonds with mean bond length equal to 1.1.}
\end{thebibliography}

\end{document}